\documentclass[letterpaper,conference]{IEEEtran}

\usepackage[T1]{fontenc}
\usepackage[nocompress]{cite}
\usepackage{graphicx}
\usepackage{booktabs}
\usepackage{array}
\usepackage{amsmath}
\usepackage{url}
\newcommand{\papertitle}{Using the Whole Slice:\\ AlphaFold2 Inference on an Eight-Chip TPU~v5e}
\newsavebox{\gentablebox}

\newcommand{\vonequote}[1]{``#1''}
\newcommand{\AFthree}{AlphaFold~3}
\makeatletter
\newcommand{\keepwithlist}{\par\penalty-100\begingroup\dimen@=5\baselineskip\dimen@ii=\pagegoal
  \advance\dimen@ii-\pagetotal\ifdim\dimen@>\dimen@ii\ifdim\dimen@ii>\z@\vfil\fi\penalty-\@M\fi\endgroup}
\makeatother

\newcommand{\capFigScaling}{Multi-query \texttt{pmap} on the matched grid. Each chip ran one query of the repeated-alphabet sequence, all eight chips were visible, and the timed region was \texttt{apply}. Each marker is one run. (a) Steady-state throughput against the number of chips, on log axes. Dashed lines show one power-law fit to all 16 points, $4528 \times \mathrm{chips}^{0.9631} \times \mathrm{length}^{-1.5724}$ queries/s, drawn at each length. (b) Parallel efficiency, the speedup over one chip divided by the number of chips. The dotted line marks 100\%. In panel (b), vertical bars at 1000 residues give the range over the rounding of the records. In panel (b), markers are offset horizontally for readability. Table~\ref{tab:grid} gives the values.}
\newcommand{\capTabPlatforms}{Baseline Configuration on Three Platforms, by Session}
\newcommand{\capTabExperiments}{Baseline Configuration and TPU Experiments}
\newcommand{\capTabMultichip}{Ways of Mapping AlphaFold2 onto the Eight-Chip Slice at 118 Residues}
\newcommand{\capTabSweeps}{Single-Chip Sweeps}
\newcommand{\capTabGrid}{Throughput, Speedup, and Parallel Efficiency on the Matched Grid}

\newif\ifappendix
\ifdefined\noappendix\appendixfalse\else\appendixtrue\fi

\begin{document}
\bstctlcite{BSTcontrol}

\title{\papertitle}

\author{\IEEEauthorblockN{Lorenzo Pazienza}
\IEEEauthorblockA{LUISS Guido Carli University\\Rome, Italy\\lorenzo.pazienza@gmail.com}
\and
\IEEEauthorblockN{Ihab El Bani}
\IEEEauthorblockA{Al Akhawayn University\\Ifrane, Morocco\\elbani.ihab@gmail.com}}

\maketitle

\begin{abstract}
AlphaFold2 is written in JAX, a framework that treats each chip of a tensor processing unit (TPU) slice as a separate device.
A JAX program does not spread its work over the chips by itself, so a user who runs AlphaFold2 on a slice can pay for chips
that do nothing. We found no published measurement of how AlphaFold2 inference uses a multi-chip TPU slice.
Our goal was to measure whether AlphaFold2 inference uses all eight chips of a TPU v5e slice and to find which way of mapping the work in JAX puts every chip to work.
All 12 experiments used random weights and a multiple sequence alignment that contains only the input sequence, so we studied execution and not prediction quality.
Ten experiments ran \texttt{model\_3} with no recycling, and the other two varied the model and the number of recycles.
By default, only one chip held data. We call each input sequence a query.
We batched up to eight queries of 118 residues on one chip with \texttt{jax.vmap}, and the throughput did not rise.
With \texttt{jax.pmap}, the eight chips ran eight independent queries in parallel.
On a grid of 1 to 8 chips and 100 to 1000 residues, the full slice gave 6.5 to 7.9 times the throughput of one chip.
Without sharding annotations, JAX's automatic partitioning did not make one query run faster. Most settings ran once,
and the logs of the TPU runs were not kept.
\end{abstract}


\section{Introduction}\label{sec:intro}

AlphaFold2 predicts the three-dimensional structure of a protein from its amino-acid sequence~\cite{jumper2021alphafold}.
The public implementation of AlphaFold2 is written in JAX~\cite{alphafoldrepo}. JAX compiles the model with the
accelerated linear algebra (XLA) compiler for central processing units (CPUs), graphics processing units (GPUs), and
tensor processing units (TPUs)~\cite{frostig2018jax}. A TPU slice connects several TPU chips, and JAX treats each chip
as a separate device. A JAX program on a slice does not spread its work over every chip by itself. The developers of
AlphaFold2 answered in GitHub issue 312 that they do not parallelize the model in JAX over several GPUs~\cite{af2issue312}.
In issue 611, the developers answered that AlphaFold2 does not support running on a TPU~\cite{af2issue611}.
The program has to map its work to the chips, for example with
\texttt{jax.pmap}~\cite{jaxpmap}. At list price, a slice costs the same whether one chip works or all eight chips work
(Section~\ref{sec:res-cost}). As a result, a user who runs AlphaFold2 on a slice without mapping the work can pay for
chips that do nothing.

Our goal was to answer two questions for a user of AlphaFold2 on a TPU slice.
First, does AlphaFold2 inference use all eight chips of the slice by default? Second, which way of mapping the work in JAX puts the other chips to work?
For that mapping, we also ask how throughput grows with the number of chips.

This study began as a project in a high-performance computing course at Stanford University. TPUs were a main topic.
Afterward, the project grew into the study that we report here.
We ran AlphaFold2 inference on an eight-chip TPU v5e slice of the course cluster in 12 experiments.
We compared four ways of mapping the work. The four ways are the default placement, batching with \texttt{jax.vmap},
multi-query mapping with \texttt{jax.pmap}, and automatic partitioning. The model used random weights, so the study
concerns execution and not the quality of the predicted structures.

Work on AlphaFold2 performance has mostly targeted GPUs.
The authors of AlphaFold2 trained the model on 128 TPU v3 cores and reported inference times on V100 GPUs~\cite{jumper2021alphafold}.
OpenFold reimplements AlphaFold2 in PyTorch~\cite{ahdritz2024openfold}. FastFold runs AlphaFold2 on GPU
clusters~\cite{cheng2024fastfold}. ParaFold splits the pipeline between CPU nodes and GPUs~\cite{zhong2022parafold}.
ParaFold sorts its input sequences by length, so that JAX compiles the model once for many of
them~\cite{zhong2022parafold}.
ColabFold speeds up batch predictions about 90-fold, in part by avoiding recompilation~\cite{colabfold2022}.
APACE runs AlphaFold2 on 200 NVIDIA A100 GPUs~\cite{apace2024}.
MassiveFold runs the many predictions of massive sampling in parallel, on systems from a single computer to a large GPU infrastructure~\cite{raouraoua2024massivefold}.
ManyFold trains a protein folding model on TPUs~\cite{villegasmorcillo2023manyfold}.
One of the 50 TPU workloads of JAXBench is the triangle multiplication of AlphaFold2~\cite{jaxbench2026}.
A concurrent working paper measures ColabFold inference on one Cloud TPU v5e or v6e chip and AlphaFold2 model
inference on one v6e chip~\cite{forsyth2026timetoscience}.
On November 23, 2026, we searched OpenAlex, DBLP, arXiv, IEEE Xplore, Crossref, Europe PMC, the Google Cloud blog, and the GitHub issues of AlphaFold2 and ColabFold.
Apart from the earlier versions of this paper\ifappendix{} (Appendix~\ref{app:changes})\fi, we found no published
measurement of AlphaFold2 inference across a multi-chip TPU slice.

In this paper, a query is one input sequence that the model predicts once. Besides the experiments on the slice, we ran
one baseline configuration (Section~\ref{sec:workload}) on one TPU v5e chip of the course cluster and, on Google Colab, on an NVIDIA T4 GPU and on a CPU runtime with two logical CPUs of an Intel Xeon processor (Section~\ref{sec:platforms}).

By default, AlphaFold2 ran on one chip of the slice, and the other seven chips held no data.
We batched up to eight queries on one chip with \texttt{jax.vmap}, and the throughput did not rise.
With \texttt{jax.pmap}, the eight chips ran eight independent queries in parallel.
On a grid of chip counts and of sequence lengths, the full slice gave 6.53 to about 7.9 times the throughput of one chip.
Without sharding annotations, JAX's automatic partitioning did not shorten the steady-state call for one query. We also
built an external ensemble, which runs differently featurized copies of one query outside the model. The external
ensemble ran on all eight chips.
When two processes shared a compilation cache, the second process took less time for the parameter initialization and for the first call (Section~\ref{sec:timing}).

We report the comparison with the T4 and with the CPU runtime as context and not as a result. The three platforms ran
in separate sessions and with different copies of the benchmark script. On the T4 and on the CPU runtime, the times
changed between sessions (Section~\ref{sec:res-platforms}). The results on the slice come from one platform and from
one period.

\keepwithlist
This paper makes three contributions.
\begin{itemize}
\item We measure how AlphaFold2 inference uses an eight-chip TPU v5e slice under the four ways of mapping the work
described above (Sections~\ref{sec:res-default} to~\ref{sec:res-auto}).
\item We give the throughput on a grid of chip counts and of sequence lengths when each chip runs one query through
\texttt{jax.pmap} (Section~\ref{sec:res-pmap}). We also give the cost of the queries at list price
(Section~\ref{sec:res-cost}).
\item We list the script, the input, the timed region, and the number of runs of every experiment. We also state
which records were not kept (Section~\ref{sec:method}).
\end{itemize}

This paper is version 4 of arXiv:2609.34818.
\ifappendix Appendix~\ref{app:changes} lists the claims that changed from the earlier versions.\fi

\section{Background}\label{sec:background}

\subsection{Chips, Slices, and Hosts}\label{sec:bg-tpu}
A TPU v5e chip has 16~GiB of high-bandwidth memory (HBM)~\cite{googletpuv5e}.
Google's TPUs link their chips with an inter-chip interconnect (ICI)~\cite{jouppi2020domain}.
The full TPU v5e configuration of 256 chips has a two-dimensional torus topology~\cite{googletpuv5e}.
Our slice had eight chips in a 2~$\times$~4 topology.
A host is the CPU machine that runs the Python process and drives the chips. JAX treats each chip as one device of that
process. In this paper, a chip is one TPU v5e chip. A slice is the set of eight chips that each Kubernetes Job received
(Section~\ref{sec:platforms}).

\subsection{Compilation and Mapping in JAX}\label{sec:bg-jax}
Before compiling a function, JAX traces the function into an intermediate program.
\texttt{jax.jit} does three things~\cite{jaxjit}. It compiles a function with XLA when the function is first called, it caches the compiled code, and it reuses the cached code in later calls.
In this paper, the word compilation covers both tracing and XLA compilation, because our measurements do not separate the two steps.
JAX can also store compiled programs on disk in a compilation cache, so that a later run can load the programs instead of compiling again~\cite{jaxcache}.

JAX offers several ways to run more than one query. \texttt{jax.vmap} vectorizes a function over a batch axis of its
inputs~\cite{jaxvmap}. By default, \texttt{jax.vmap} runs the whole batch on one device.
\texttt{jax.pmap} replicates
a function and runs each replica on its own device. Each device receives one element of the leading axis of the
input~\cite{jaxpmap}. The documentation of \texttt{jax.pmap} now recommends \texttt{jax.shard\_map}
instead~\cite{jaxpmap}.
Since JAX 0.10.0, \texttt{jax.pmap} is implemented as \texttt{jax.jit} applied to \texttt{jax.shard\_map}~\cite{jaxchangelog}.
Our external ensemble (Section~\ref{sec:res-auto}) uses \texttt{jax.lax.pmean}, which computes the mean of an array over
the devices of a mapped axis~\cite{jaxpmean}.

JAX can also partition one computation automatically over a mesh. A mesh is a grid of devices with named axes. The
GSPMD partitioner of XLA starts from the sharding annotations that the program places on some arrays. The partitioner
derives the placement of the other arrays from these annotations~\cite{xu2021gspmd}. The OpenXLA guide to Shardy for
JAX users says that JAX now always uses the Shardy partitioner~\cite{shardy2026}.
We read this guide on November 23, 2026.
According to the JAX change log, Shardy became the default partitioner in JAX 0.7.0, and JAX 0.12.1 removed the GSPMD partitioner~\cite{jaxchangelog}.
The committed Jobs installed JAX 0.10.2 and set no partitioner flag, so Shardy was the default.
The runs did not record which partitioner ran. In Shardy, an array is replicated along each mesh axis that does not
shard one of the dimensions of the array~\cite{shardyrepr}. This rule matters for our partitioning experiment, which
placed no sharding annotations (Section~\ref{sec:res-auto}).

\subsection{AlphaFold2 Inference Pipeline}\label{sec:bg-af2}
AlphaFold2 reads a multiple sequence alignment (MSA). An MSA lines up the query with related protein sequences. A
stack of Evoformer blocks updates a representation of the MSA and a representation of residue pairs. Residues are the
amino acids of the protein chain. The model also reads extra MSA rows, a second and larger set of aligned sequences.
Before the 48 Evoformer blocks, a separate stack of four smaller blocks processes these extra rows~\cite{alphafoldrepo}.
A structure module then turns these representations into atom coordinates~\cite{jumper2021alphafold}. The public
implementation uses Haiku~\cite{haiku2020github}. Haiku is a neural-network library for JAX. The model can recycle its
output as a new input for a fixed number of iterations. The public implementation runs recycling in a
\texttt{while\_loop}. By default, the monomer presets of the public implementation run three recycling iterations and
predict each target with five models, \texttt{model\_1} to \texttt{model\_5}~\cite{alphafoldrepo}.

The implementation can also run an ensemble. The members of an ensemble are differently featurized copies of the same
input. Featurization turns the sequence and the MSA into the arrays that the model reads. The implementation computes the
first member outside a loop and the others in a sequential Haiku \texttt{while\_loop}~\cite{alphafoldrepo}.
The authors of AlphaFold2 turn off ensembling for most inference~\cite{jumper2021alphafold}. In the released code, the
preset \texttt{monomer\_casp14} uses eight ensemble members, and the other presets use one~\cite{alphafoldrepo}.
Our runs used one ensemble member, so the ensemble \texttt{while\_loop} ran no iteration in the timed calls.

Our scripts time one of two methods of the model. The \texttt{apply} method is the forward pass on the device. The
\texttt{predict} method runs \texttt{apply} and then computes confidence metrics on the host with NumPy.

\section{Experimental Setup}\label{sec:method}

\subsection{Platforms, Sessions, and Software}\label{sec:platforms}
\begin{table*}[t]
\caption{\capTabPlatforms}\label{tab:platforms}
\centering\small\sbox{\gentablebox}{\begin{tabular*}{\linewidth}{@{\extracolsep{\fill}}lllcrrrr@{}}
\toprule
Platform & Session (UTC) & Script & JAX & Calls & \begin{tabular}[b]{@{}r@{}}Parameter\\initialization (s)\end{tabular} & \begin{tabular}[b]{@{}r@{}}First call\\(s)\end{tabular} & \begin{tabular}[b]{@{}r@{}}Steady-state\\call (s)\end{tabular} \\
\midrule
TPU v5e, one chip & by Aug. 8, 2026, 17:23 & not identified & not recorded & 1 & 36.60 & 27.78\dag & 0.470 \\
T4 & Aug. 8, 2026, 04:50 & notebook copy & not recorded & 1 & 109.16 & -- & 13.086 \\
Xeon CPU runtime & Aug. 8, 2026, 05:19 & notebook copy & not recorded & 1 & 41.99 & -- & 212.113 \\
\midrule
T4 & Sept. 15, 2026, 19:01 & 86fca03 & 0.10.2 & 5 & 102.60 & 39.59 & 6.57 $\pm$ 0.06 \\
Xeon CPU runtime & Sept. 15, 2026, 18:04 & 86fca03 & 0.10.2 & 5 & 59.02 & 385.48 & 348.9 $\pm$ 4.7 \\
Xeon CPU runtime & Sept. 16, 2026, 12:37 & 9e1a07e & 0.10.2 & 5 & 59.87 & 373.24 & 351.9 $\pm$ 4.3 \\
Xeon CPU runtime & Sept. 17, 2026, 08:52 & 9e1a07e & 0.10.2 & 5 & 60.30 & 394.69 & 359.3 $\pm$ 2.8 \\
\bottomrule
\end{tabular*}%
}\usebox{\gentablebox}\par\vspace{2pt}\begin{minipage}[t]{\wd\gentablebox}\footnotesize\leftskip=0pt\rightskip=0pt
  \parfillskip=0pt plus 1filAll runs use the baseline configuration (Section~\ref{sec:workload}). The CPU runtime had two logical CPUs of an Intel Xeon processor. Calls is the number of steady-state calls. With more than one call, the steady-state call is given as a mean $\pm$ sample standard deviation. Script gives the commit of the benchmark script. For the TPU, the Session column gives the time of the commit that holds the TPU result. For the T4 and for the CPU runtime, the Session column gives the start of the session, which is the first time stamp in the log output of the notebook. A dash marks the first calls of August 8, 2026, on the T4 and on the CPU runtime. These first calls ran with the profiler. The TPU has no run without the profiler, so the table shows its first call. \dag{} The first call on the TPU probably includes the profiler teardown (Section~\ref{sec:res-first}). Single times keep the decimals that the scripts round to, including trailing zeros.
\end{minipage}\par
\end{table*}

A platform is one of our three hardware targets, which are the TPU v5e chip of the course cluster, the NVIDIA T4 GPU on Google Colab, and the Google Colab CPU runtime with two logical CPUs of an Intel Xeon processor.
A session is one Colab runtime or one Kubernetes Job in which a set of runs took place.
Table~\ref{tab:platforms} lists the stand-alone baseline run of each session.

The TPU runs used a Google Kubernetes Engine cluster of the course, in the us-west4 region. Each Kubernetes Job
requested the whole slice of eight chips. Some Jobs ran several experiments.
The Job definitions in the repository start from a \texttt{python:3.12-slim} image and install \texttt{jax[tpu]==0.10.2}.
The Job definitions pin no other package and set a lower bound only for \texttt{tensorflow-cpu}.
The Job definitions clone AlphaFold2 from its default branch, and the runs did not record the commit of AlphaFold2.
 The TPU runs did not record their dates.
Most TPU results are in the first commit of our benchmark repository. This commit is dated August 8, 2026, 17:23 UTC.
We committed the results of the matched grid (Section~\ref{sec:res-pmap}) on August 8, 2026, at 18:05 UTC. We committed the results of the external ensemble on August 11, 2026, at 22:38 UTC.

The baseline runs on the T4 and on the CPU runtime took place in two Colab sessions on August 8, 2026, at 04:50 UTC and 05:19 UTC, respectively.
These runs used a copy of the benchmark script inside the notebooks. We call this copy the notebook copy.
The notebooks in commit 2755250 hold this copy and the output of the runs. This copy differs from the scripts in \texttt{src/}.
The first calls of these two Colab runs used the profiler. Table~\ref{tab:platforms} does not show these first calls.
The two Colab sessions of August 8, 2026, did not record their JAX version or their host CPU.
The file \texttt{results/comparison.md} of the repository on August 12, 2026 (commit 612c5a2, 17:37 UTC) describes the CPU runtime as Colab's standard allocation of an Intel Xeon with two vCPUs.
We cannot identify the script of the baseline run on the TPU.
The result file of the TPU baseline run lacks the field \texttt{host\_processor}, which the notebook copy writes. The result file also contains none of the extra fields that the committed script writes.
The format of the file is closer to the notebook copy than to the committed script.

From September 15 to 17, 2026, we ran the committed script on the T4 and on the CPU runtime.
The CPU runtime ran in three sessions, on September 15, 2026, at 18:04 UTC, on September 16, 2026, at 12:37 UTC, and on September 17, 2026, at 08:52 UTC.
The T4 ran in one session, on September 15, 2026, at 19:01 UTC. Table~\ref{tab:platforms} explains these times.
In the three CPU sessions, the CPU runtime had an Intel Xeon processor at 2.20~GHz with two logical CPUs.
Each session ran the script twice. One run used the profiler, and the other did not. For these four sessions, tables
and text report the runs without the profiler. Section~\ref{sec:res-first} also uses the runs with the profiler. Each
of the four sessions recorded its commit, its package versions, and its host CPU.
These sessions ran commits 86fca03 (September 15, 2026, 17:57 UTC) and 9e1a07e (September 16, 2026, 12:35 UTC), which have the same benchmark scripts.
All four September sessions ran JAX 0.10.2 and AlphaFold2 at commit c77e5d2~\cite{alphafoldrepo}.
We did not repeat the TPU runs. Our access to the cluster ended with the course.
In the folder \texttt{cloud/}, we prepared a repeat of the TPU runs on Cloud TPU. We had no cloud resources when we
prepared the repeat, and we did not run it later.

\subsection{Workload and Inputs}\label{sec:workload}
\begin{table*}[t]
\caption{\capTabExperiments}\label{tab:experiments}
\centering\small\setlength{\tabcolsep}{3.5pt}\sbox{\gentablebox}{\begin{tabular}{@{}rlllllrl@{}}
\toprule
\# & Experiment & Varied & Script & Input & Timed & Runs & File \\
\midrule
0 & Baseline configuration & platform & -- & fixed, 118 & predict & 1 & \texttt{result\_*}, \texttt{repro/} \\
\midrule
1 & Sequence length & 60, 120, 250, 500 residues & single & alphabet & predict & 1 & \texttt{sequence\_length} \\
2 & Recycling & 0, 1, 3 recycles & single & fixed, 118 & predict & 1 & \texttt{recycle\_depth} \\
3 & Chip visibility & 1, 8 chips visible & single & fixed, 118 & predict & 1 & \texttt{chip\_visibility} \\
4 & Parameter type & float32, bfloat16 & single & fixed, 118 & predict & 1 & \texttt{precision} \\
5 & Batching with \texttt{vmap} & $B$ = 1, 2, 4, 8 & vmap & alphabet, 118 & apply & 1 & \texttt{batching} \\
6 & Compilation cache & empty or filled cache & single & fixed, 118 & predict & 1 & \texttt{compilation\_cache} \\
7 & Models & \texttt{model\_3}, \texttt{model\_4}, \texttt{model\_5} & single & fixed, 118 & predict & 1 & \texttt{model\_comparison} \\
8 & Repeated runs & none & single & fixed, 118 & predict & 3 & \texttt{repeated\_runs} \\
9 & Multi-query \texttt{pmap} & 8 queries on 8 chips & pmap & alphabet, 118 & apply & 1 & \texttt{sharding} \\
10 & Automatic partitioning & mesh of 8 chips & partitioning & fixed, 118 & predict & 2 & \texttt{sharding} \\
11 & External ensemble & 8 seeds on 8 chips & ensemble & fixed, 118 & apply, pmean & 1 & \texttt{ensemble\_shard} \\
12 & Matched grid & \begin{tabular}[t]{@{}l@{}}1, 2, 4, 8 chips $\times$\\100, 250, 500, 1000 residues\end{tabular} & pmap & alphabet & apply & 1 & \texttt{scaling\_law\_data} \\
\bottomrule
\end{tabular}%
}\usebox{\gentablebox}\par\vspace{2pt}\begin{minipage}[t]{\wd\gentablebox}\footnotesize\leftskip=0pt\rightskip=0pt
  \parfillskip=0pt plus 1filRuns counts the runs per setting on the TPU. Experiment 0 is the stand-alone baseline run of each session in Table~\ref{tab:platforms}, which also gives its script. Each session of September 15 to 17, 2026, also ran the baseline configuration once with the profiler. On the TPU, the baseline configuration is also one setting of experiments 2, 4, and 7. Both settings of experiments 3 and 6 use the baseline configuration. All three runs of experiment 8 use the baseline configuration. With experiment 0, the TPU runs of the baseline configuration are the 11 runs of Section~\ref{sec:experiments}. The scripts are in \texttt{src/}. \emph{Single} is \texttt{spike\_tpu\_forward\_pass.py}, \emph{vmap} is \texttt{spike\_batch\_forward\_pass.py}, \emph{pmap} is \texttt{spike\_pmap\_forward\_pass.py}, \emph{partitioning} is \texttt{spike\_meshshard\_forward\_pass.py}, and \emph{ensemble} is \texttt{spike\_ensemble\_shard\_forward\_pass.py}. \emph{Alphabet} and \emph{fixed} are the repeated-alphabet and fixed sequences, followed by the length in residues when the experiment does not vary it. File gives the start of the file name. Files are in \texttt{results/sweep/}. The files of experiment 0 are in \texttt{results/}. Unless an experiment varies one of these settings, all runs use \texttt{model\_3}, zero recycles, and float32 parameters.
\end{minipage}\par
\end{table*}

Every run, except those of the model comparison, used \texttt{model\_3} of AlphaFold2.
This model needs no templates. Templates
are known structures of related proteins that \texttt{model\_1} and \texttt{model\_2} take as input.
The script built the model with random weights. Haiku creates the random weights in the parameter initialization call.
 The
script never loaded the trained checkpoint. The MSA held only the query sequence, and no run searched a sequence
database. The featurization pads the MSA features to a fixed number of rows.
With one sequence, \texttt{model\_3} still runs on 512 MSA rows and 5120 extra MSA rows~\cite{alphafoldrepo}.
Only the first MSA row holds the query, and the other rows are padding.
Every run used zero recycles, one ensemble member, float32 parameters, and random seed 0, unless an experiment varied one of these settings.
Seed 0 sets both the parameter initialization and the featurization. In the external ensemble, only the featurization seed varies, from 0 to 7.
AlphaFold2's source ran unmodified. Newer JAX releases renamed the arguments \texttt{a\_min} and \texttt{a\_max} of
\texttt{jax.numpy.clip} to \texttt{min} and \texttt{max}. A small wrapper in each script renamed these arguments.

The experiments used two input families. The fixed sequence has 118 residues.
The repeated-alphabet sequence cycles through the 20 amino-acid letters up to the target length.
In runs with several queries, each query starts one letter later in the alphabet. Table~\ref{tab:experiments} gives
the family of each experiment. The baseline configuration is the single-query script (Section~\ref{sec:experiments})
on the fixed sequence with \texttt{model\_3}, zero recycles, and float32 parameters.
The baseline configuration times \texttt{predict}.

\subsection{Timing, Memory, and Metrics}\label{sec:timing}
The scripts time each call of the model with \texttt{time.time} on the host. They wait for the outputs of the first
call and of the steady-state calls with \texttt{jax.block\_until\_ready}. Parameter initialization is the
\texttt{init\_params} call, which traces the model and creates the random weights. The first call is the earliest call
of \texttt{predict} or \texttt{apply} at a new input shape. The first call traces, compiles, and executes the model. A
steady-state call is a later call at the same input shape. A steady-state call reuses the compiled program.
 The single-query
and automatic-partitioning scripts time the parameter initialization, the first call, and the steady-state calls.
The
\texttt{vmap}, \texttt{pmap}, and ensemble scripts time the first call and one steady-state call. The Colab runs of August 8, 2026,
and all TPU runs timed one steady-state call, and the runs of September 15 to 17, 2026, without the profiler timed five.

The scripts time two different regions. The single-query script and the automatic-partitioning script time
\texttt{predict}, which includes the confidence metrics on the host. The \texttt{vmap}, \texttt{pmap}, and ensemble
scripts time \texttt{apply} (Table~\ref{tab:experiments}).
The ensemble script runs \texttt{jax.lax.pmean} inside the same \texttt{jax.pmap} call as \texttt{apply}, so the timed region of the ensemble script also contains this collective operation.
 Except in the runs of September 15 to 17, 2026, without the profiler, the
single-query script and the \texttt{vmap} script run the first call inside \texttt{jax.profiler.trace}. This context
records a profile of the call. The profiler teardown is the time that the \texttt{jax.profiler.trace} context takes to
close after the first call returns. The teardown lies inside the timer of the first call.

Throughput is the number of queries that steady-state calls complete per second. We report throughput in queries/s.
The speedup is the throughput on $n$ chips divided by the throughput on one chip, with the same script, input family,
sequence length, and timed region. With multi-query \texttt{pmap}, each chip runs one query, so the speedup is a
weak-scaling measure of throughput and not a shorter time for one query. The one-chip reference of this speedup is
\texttt{pmap} on one chip. The parallel efficiency is the speedup divided by $n$. On one chip, multi-query
\texttt{pmap} runs one query. Every \texttt{pmap} run had all eight chips visible. We did not measure the utilization
of a chip. In this paper, a chip that works is a chip that holds data and runs a replica.

HBM in use is the value of \texttt{bytes\_in\_use} that JAX reports for a chip after the first steady-state call. This
value is the device memory allocated right after the first steady-state call. Peak memory is the value of
\texttt{peak\_bytes\_in\_use} that JAX reports for a chip at the same moment. We did not check whether this value
includes the temporary buffers of compiled programs.

\subsection{Experiments}\label{sec:experiments}
Table~\ref{tab:experiments} lists the baseline configuration and the 12 TPU experiments, with script, input, timed region, and number of runs.
Experiments in one Job ran one after another in the same container, in a fixed order. Most settings ran once. The
repeated-runs experiment ran the baseline configuration three times on the TPU, in one Job.
Across these three runs, the coefficient of variation was 0.12\% for the steady-state call, 0.82\% for the parameter initialization, and 1.00\% for the first call.
We computed the coefficient of variation with the sample standard deviation.
The steady-state times are rounded to 1~ms, about 0.2\% of the call, so the resolution of the records limits this coefficient.
We ran the baseline configuration 11 times on the TPU. Ten of these runs started with an empty compilation cache.
The note of Table~\ref{tab:experiments} lists the settings that make up these runs. We also tried to make only two or
four chips visible to the single-query script. The file \texttt{results/sweep/chip\_visibility.md} records that the
runs with two or four visible chips stopped with an error. We have no log of these runs.

\subsection{Provenance of the Result Files}\label{sec:records}
Most TPU results were committed together with their scripts. The files in \texttt{results/sweep/} are summaries of the
Job output. These summaries are not the files that the scripts wrote. Most field names in these summaries differ from
those that the scripts write.
Every memory value in the summaries is a whole number of megabytes.
The logs of the Jobs are not in the repository, and we have not found these logs. The first author produced these
summaries from the output of the Jobs with Claude (Anthropic). We have no record of how each value was transferred. We
report the values in these summaries unchanged. We checked the summaries against each other. In the \texttt{vmap} and
\texttt{pmap} summaries, each throughput equals the number of queries divided by the duration of the steady-state call,
within the rounding of that duration.
The 11 single-query runs of the baseline configuration come from seven files. These runs took 0.469~s to 0.471~s.
These checks cannot replace the output of the Jobs.

\section{Results}\label{sec:results}

\subsection{Default Placement and Batching on One Chip}\label{sec:res-default}
\begin{table*}[t]
\caption{\capTabMultichip}\label{tab:multichip}
\centering\small\setlength{\tabcolsep}{4pt}\sbox{\gentablebox}{\begin{tabular*}{\linewidth}{@{\extracolsep{\fill}}lllccrlrr@{}}
\toprule
Configuration & Input & Timed & \begin{tabular}[b]{@{}c@{}}Queries\\per call\end{tabular} & \begin{tabular}[b]{@{}c@{}}Chips\\with data\end{tabular} & \begin{tabular}[b]{@{}r@{}}HBM, chip 0\\(MB)\end{tabular} & \begin{tabular}[b]{@{}l@{}}HBM, chips\\1--7 (MB)\end{tabular} & \begin{tabular}[b]{@{}r@{}}Steady-state\\call (s)\end{tabular} & \begin{tabular}[b]{@{}r@{}}Throughput\\(queries/s)\end{tabular} \\
\midrule
Default & fixed & predict & 1 & 1 of 8 & 463 & 0 & 0.470 & -- \\
\texttt{vmap}, $B$ = 1 & alphabet & apply & 1 & not kept & 487 & not kept & 0.4875 & 2.051 \\
\texttt{vmap}, $B$ = 2 & alphabet & apply & 2 & not kept & 532 & not kept & 1.0374 & 1.928 \\
\texttt{vmap}, $B$ = 4 & alphabet & apply & 4 & not kept & 630 & not kept & 2.6905 & 1.487 \\
\texttt{vmap}, $B$ = 8 & alphabet & apply & 8 & not kept & 821 & not kept & 5.3053 & 1.508 \\
Multi-query \texttt{pmap} & alphabet & apply & 8 & 8 of 8 & 644 & 445--469 & 0.5435 & 14.718 \\
Automatic partitioning & fixed & predict & 1 & not kept & \multicolumn{2}{c}{463} & 0.472, 0.473 & -- \\
External ensemble & fixed & apply, pmean & 1 & 8 of 8 & 624 & 427--450 & 0.5381 & -- \\
\bottomrule
\end{tabular*}%
}\usebox{\gentablebox}\par\vspace{2pt}\begin{minipage}[t]{\wd\gentablebox}\footnotesize\leftskip=0pt\rightskip=0pt
  \parfillskip=0pt plus 1filAll runs use 118 residues on a slice with eight chips visible. The scripts compute throughput as the queries per call divided by the unrounded steady-state call. We give no throughput for the rows that time \texttt{predict} or for the external ensemble, which computes one query. Chips with data counts the chips with nonzero HBM in use. Not kept marks a value that is missing from the summary. For automatic partitioning, each run kept one value of HBM in use, 463~MB, and the summary does not say whether 463~MB is the mean over the eight chips or the HBM in use on chip~0 alone. MB is 1,000,000 bytes. Rows differ in script, so ratios between rows are not speedups. \emph{Alphabet} and \emph{fixed} are the repeated-alphabet and fixed sequences. Times keep the decimals that the scripts round to, including trailing zeros.
\end{minipage}\par
\end{table*}

By default, AlphaFold2 inference ran on one chip of the slice. We ran the single-query script with one chip visible and
with all eight chips visible.
The steady-state call took 0.469~s with one chip visible and 0.470~s with eight chips visible.
With eight chips visible, chip~0 had 463~MB of HBM in use, and the other seven chips had none (Table~\ref{tab:multichip}).
The run gave no sign that the other chips took part.

Batching with \texttt{vmap} runs $B$ queries in one call on one chip. Batching did not raise throughput at any batch size
we tried.
With $B = 1$, 2, 4, and 8, the throughput was 2.051~queries/s, 1.928~queries/s, 1.487~queries/s, and 1.508~queries/s, respectively.
From the recorded times, the three larger batches reached 0.94, 0.72, and 0.74 times the throughput of $B$ = 1.
With $B$ = 4 and with $B$ = 8, one call took 38\% and 36\% longer than $B$ calls with $B$ = 1 would take. We did not study the cause.
The HBM in use on chip~0 grew from 487~MB to 821~MB. The script recorded every chip, but the summary of the \texttt{vmap} runs kept chip~0 only.

\subsection{Multi-Query pmap and the Matched Grid}\label{sec:res-pmap}
Multi-query \texttt{pmap} runs one independent query on each chip through \texttt{jax.pmap}.
Multi-query \texttt{pmap} ran eight queries of 118 residues in parallel on all eight chips of the slice.
Chip~0 had 644~MB of HBM in use, and the other seven chips had 445~MB to 469~MB.
The steady-state call took 0.5435~s, and the throughput was 14.718~queries/s.
Memory alone does not show how the work was divided, but \texttt{jax.pmap} runs one replica on each device~\cite{jaxpmap}, so each chip ran one query.

\begin{figure}[t]
\centering
\includegraphics{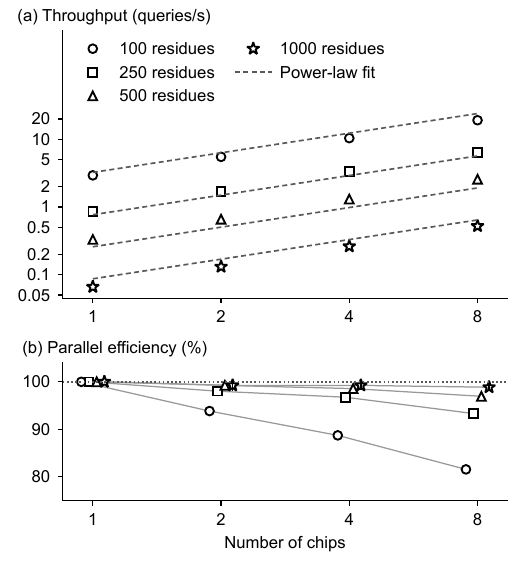}
\caption{\capFigScaling}
\label{fig:scaling}
\end{figure}

\begin{table*}[t]
\caption{\capTabGrid}\label{tab:grid}
\centering\small\sbox{\gentablebox}{\begin{tabular*}{\linewidth}{@{\extracolsep{\fill}}rrrrrccc@{}}
\toprule
& \multicolumn{4}{c}{Throughput (queries/s)} & & & \\ \cmidrule(lr){2-5}
Residues & 1 chip & 2 chips & 4 chips & 8 chips & \begin{tabular}[b]{@{}c@{}}Speedup on\\8 chips\end{tabular} & \begin{tabular}[b]{@{}c@{}}Parallel efficiency\\on 8 chips (\%)\end{tabular} & \begin{tabular}[b]{@{}c@{}}Extra time per call\\on 8 chips (s)\end{tabular} \\
\midrule
100 & 2.935 & 5.509 & 10.417 & 19.153 & 6.53 & 81.6 & 0.077 \\
250 & 0.864 & 1.694 & 3.344 & 6.450 & 7.47 & 93.3 & 0.083 \\
500 & 0.332 & 0.659 & 1.309 & 2.575 & 7.76 & 97.0 & 0.095 \\
1000 & 0.066 & 0.131 & 0.262 & 0.522 & 7.84--7.98 & 98.0--99.8 & 0.04--0.31 \\
\bottomrule
\end{tabular*}%
}\usebox{\gentablebox}\par\vspace{2pt}\begin{minipage}[t]{\wd\gentablebox}\footnotesize\leftskip=0pt\rightskip=0pt
  \parfillskip=0pt plus 1filEach chip ran one query of the repeated-alphabet sequence with multi-query \texttt{pmap}, all eight chips were visible, and the timed region was \texttt{apply}. Each point ran once. The extra time is the steady-state call on eight chips minus the call on one chip, both derived from the recorded throughputs. The summary gives the one-chip throughput at 1000 residues with two digits, so the last three values at this length are ranges over the rounding of the records. We round the ends of each range outward.
\end{minipage}\par
\end{table*}

The matched grid measures the speedup with the same script, input family, and timed region on every point.
The matched grid ran multi-query \texttt{pmap} on 1, 2, 4, and 8 chips (Fig.~\ref{fig:scaling}). The sequence lengths were 100~residues, 250~residues, 500~residues, and 1000~residues.
Each of the 16 points ran once.
The Job definition committed with these results runs the 16 points in one Job, one process per point, in the order 1, 2, 4, and 8 chips, each at 100, 250, 500, and 1000 residues.
Table~\ref{tab:grid} gives the 16 throughputs. Throughput grew with the number of chips.
At 100 residues, throughput rose from 2.935~queries/s on one chip to 19.153~queries/s on eight chips.
On eight chips, the speedup was 6.53 at 100 residues, 7.47 at 250 residues, 7.76 at 500 residues, and 7.84 to 7.98 at 1000 residues.
At the same four lengths, the parallel efficiency was 81.6\%, 93.3\%, 97.0\%, and 98.0\% to 99.8\%.
We computed the parallel efficiency from the unrounded speedup.
At 1000 residues, the summary gives the one-chip throughput only as 0.066~queries/s, with two digits, so we give ranges at this length.
We round the ends of each range outward.

The parallel efficiency was lowest at 100 residues. To examine this difference, we compared the time of one call on
eight chips and on one chip. The call on eight chips ran eight queries, and the call on one chip ran one. We derived
the time of each call from the recorded throughput.
At 100 residues, 250 residues, and 500 residues, the call on eight chips took 0.077~s, 0.083~s, and 0.095~s longer than the call on one chip.
Each value comes from one run on each chip count.
At 100 residues, this extra time was 23\% of the one-chip call.
At 1000 residues, the extra time lies between 0.04~s and 0.31~s. This range is too wide for a comparison with the extra time at the other lengths.
At every call, the script passes the parameters to \texttt{jax.pmap} as an unmapped argument. The script also stacks
the inputs of all queries on chip~0 before the timer. As a result, at every call on several chips, the parameters and
the inputs of every query except the first move from chip~0 to the other chips inside the timed region. The copy of the
parameters adds time that does not depend on length, and the copy of the inputs adds time that grows with length. We
did not measure these copies. Each extra time is the difference of two single runs, so these values do not show how
the extra time depends on length.

A power-law fit summarizes the grid.
We fit log throughput on log chips and log length by least squares, over all 16 points.
The fit gives a throughput of $4528 \times \mathrm{chips}^{0.9631} \times \mathrm{length}^{-1.5724}$ queries/s, with $R^2 = 0.981$ in log space.
When we fit each length separately, the chip exponent ranged from 0.904 at 100 residues to 0.995 at 1000 residues.
The fit is descriptive, and the error of the fit changes sign with length.
The fitted value exceeds the measured throughput by up to 25.5\% at 100 residues and by up to 31.6\% at 1000 residues.
The fitted value falls below the measured throughput by up to 12.7\% at 250 residues and by up to 25.7\% at 500 residues.
All four percentages are relative to the measured throughput. We do not use the fit outside the grid.

\subsection{Automatic Partitioning and External Ensemble}\label{sec:res-auto}
Automatic partitioning did not shorten the call for one query.
The two steady-state calls of this experiment took 0.472~s and 0.473~s. Single-query runs on one chip took 0.469~s to 0.471~s.
Automatic partitioning runs one query inside a JAX mesh of eight chips. The axes of this mesh are marked automatic.
Our script placed no sharding annotations and did not shard the inputs.
We searched every file of AlphaFold2 at commit c77e5d2 for \texttt{jax.sharding}, \texttt{Mesh}, \texttt{PartitionSpec}, \texttt{NamedSharding}, \texttt{with\_sharding\_\allowbreak constraint}, \texttt{pjit}, \texttt{shard\_map}, and \texttt{pmap}~\cite{alphafoldrepo}.
None of these names occurs. The TPU Jobs did not record their AlphaFold2 commit, so the code that ran may differ from
c77e5d2.

\looseness=-1 Each of the two runs kept one value of HBM in use, 463~MB.
The summary does not say whether 463~MB is the mean over the eight chips or the HBM in use on chip~0 alone.
The runs recorded neither which partitioner ran nor the placement of any array. We expect the call to run on one chip or
as a full copy on every chip. Our records cannot tell which of these two cases occurred, and we did not run a test to
find out. In both cases, no chip shares the work of another. This experiment does not test what the partitioner can do
with annotations.

AlphaFold2's own ensemble runs its members in a sequential loop (Section~\ref{sec:bg-af2}).
We also built and ran an external ensemble.
Unlike the ensemble of AlphaFold2, which combines its members inside the model, the external ensemble runs
eight full forward passes and averages one output.
We featurized one query with the eight random seeds 0 to 7. We ran one featurization per chip through \texttt{jax.pmap}.
Then \texttt{jax.lax.pmean} averaged the logits of \texttt{predicted\_lddt} over the chips~\cite{jaxpmean}. With the
random weights of \texttt{init\_params}, the \texttt{predicted\_lddt} logits are zero.
 The global configuration of AlphaFold2 sets
\texttt{zero\_init}, so the final layer of this head starts with zero weights and biases~\cite{alphafoldrepo}. The
experiment times the collective operation and not an average of informative values.
Each of the eight chips had between 427~MB and 624~MB of HBM in use.
The steady-state call took 0.5381~s.
We did not run a sequential ensemble of eight members, so we give no speedup for this experiment. The members share one
MSA and differ only by seed.
With an MSA of one sequence, the seed changes only the random masking of the MSA row. The featurization masks about 15\% of the positions of this row~\cite{alphafoldrepo}.
The experiment shows that an ensemble can run on all chips of the slice. This experiment does not measure an ensemble
that a user would run for prediction.

\subsection{First Call and Compilation Cache}\label{sec:res-first}
On the TPU, the baseline run took 36.60~s for the parameter initialization, 27.78~s for the first call, and 0.470~s for the steady-state call.
Over the ten single-query runs of this configuration with an empty compilation cache, the first call took 27.37~s to 28.80~s.
In every August version that we have, the single-query script runs the first call with the profiler. Later versions do
so by default. The first calls of these August scripts include the profiler teardown, whose duration on the TPU was not
recorded. We cannot identify the script of the baseline run (Section~\ref{sec:platforms}). The first call of the
baseline run probably includes the profiler teardown too.
In the sessions of September 15 to 17, 2026, the runs with the profiler measured the teardown on the T4 and on the CPU runtime. On the CPU runtime, the teardown took 31.7~s to 31.9~s.
On the T4, the teardown took 23.7~s.
The automatic-partitioning script runs no profiler. The first calls of this script took 14.76~s and 14.46~s.
These calls ran inside a mesh, so we do not use them to estimate the profiler teardown.

The runs of September 15 to 17, 2026, without the profiler give first calls that exclude the teardown.
On the CPU runtime, the first call took 373.2~s to 394.7~s in the three sessions, against mean steady-state calls of 348.9~s to 359.3~s.
On the T4, the first call took 39.59~s, against a mean steady-state call of 6.57~s.

On the TPU, the compilation cache shortened the parameter initialization and the first call of a second process.
We ran the baseline configuration in two processes that shared one cache directory.
Parameter initialization took 37.68~s in the first process and 5.53~s in the second. The ratio is 6.81.
The first call took 28.80~s in the first process and 15.19~s in the second. The ratio is 1.90.
Parameter initialization and the first call together took 66.48~s in the first process and 20.72~s in the second.
The steady-state call took 0.470~s in the first process and 0.469~s in the second.
Both first calls include the profiler teardown, so we do not attribute the 15.19~s of the second process to compilation.
One pair of processes ran. With the trained weights, a user loads the parameters instead of calling
\texttt{init\_params}. We did not time the loading of the trained parameters.

\subsection{Single-Chip Sweeps}\label{sec:res-sweeps}
\begin{table}[t]
\caption{\capTabSweeps}\label{tab:sweeps}
\centering\small\sbox{\gentablebox}{\begin{tabular*}{\linewidth}{@{\extracolsep{\fill}}llrr@{}}
\toprule
Sweep & Setting & \begin{tabular}[b]{@{}r@{}}First call\\(s)\end{tabular} & \begin{tabular}[b]{@{}r@{}}Steady-state\\call (s)\end{tabular} \\
\midrule
Sequence length & 60 residues & 25.95 & 0.208 \\
 & 120 residues & 24.88 & 0.408 \\
 & 250 residues & 29.37 & 1.170 \\
 & 500 residues & 31.46 & 3.031 \\
\midrule
Recycling & 0 recycles & 27.66 & 0.469 \\
 & 1 recycle & 55.46 & 0.933 \\
 & 3 recycles & 54.94 & 1.845 \\
\midrule
Parameter type & float32 & 27.88 & 0.470 \\
 & bfloat16 & 26.39 & 0.469 \\
\midrule
Models & \texttt{model\_3} & 28.72 & 0.471 \\
 & \texttt{model\_4} & 28.24 & 0.472 \\
 & \texttt{model\_5} & 28.00 & 0.403 \\
\bottomrule
\end{tabular*}%
}\usebox{\gentablebox}\par\vspace{2pt}\begin{minipage}[t]{\wd\gentablebox}\footnotesize\leftskip=0pt\rightskip=0pt
  \parfillskip=0pt plus 1filEach setting ran once on chip~0, and the script timed \texttt{predict}. The committed Job definitions of these sweeps make all eight chips of the slice visible. The sequence-length sweep uses the repeated-alphabet sequence, and the other sweeps use the fixed sequence. The rows 0 recycles, float32, and \texttt{model\_3} are three separate runs of the baseline configuration. First calls include the profiler teardown, whose duration on the TPU was not recorded (Section~\ref{sec:res-first}). Times keep the decimals that the script rounds to, including trailing zeros.
\end{minipage}\par
\end{table}

Table~\ref{tab:sweeps} gives four sweeps on one chip, and each setting in these sweeps ran once.

Between 60 and 500 residues, the steady-state call grew from 0.208~s to 3.031~s.
A least-squares fit of log time on log length over these four points gives a slope of 1.28.
Section~\ref{sec:res-pmap} fits a power law to the 16 points of the matched grid, and the length exponent of that power law is $-1.5724$. The slope of 1.28 and the exponent of $-1.5724$ cannot be compared directly. The slope describes time with \texttt{predict} over 60 to 500 residues, and the exponent describes throughput with \texttt{apply} over 100 to 1000 residues.
Between neighboring lengths, the slope of log time on log length ranged from 1.33 to 2.33 on the one-chip points of the matched grid and from 0.97 to 1.44 in the sequence-length sweep.
No single power law fits these data closely.

The steady-state call took 0.469~s, 0.933~s, and 1.845~s with 0, 1, and 3 recycles. Each recycle added 0.459~s on average.
The first call roughly doubled with one or three recycles. We did not study the cause.

The bfloat16 setting casts the parameters to bfloat16. The script does not change the activations.
The steady-state call took 0.469~s with bfloat16 parameters and 0.470~s with float32 parameters.
The HBM in use fell by 40.2\%, from 463~MB to 277~MB.
The peak memory rose by 18.5\%, from 486~MB to 576~MB.
During the cast, the chip holds both the float32 and the bfloat16 parameters, which may explain the rise of the peak memory.
 We did not
test this hypothesis. On the TPU, JAX by default computes float32 matrix products in
bfloat16~\cite{jaxmatmulprecision}. We did not change this default, which may explain why the bfloat16 parameters did
not shorten the call. We did not test this explanation.

Among the three models without templates, \texttt{model\_5} had the shortest steady-state call.
The call of \texttt{model\_5} took 0.403~s, which is 14.4\% less than the 0.471~s of \texttt{model\_3}. The call of \texttt{model\_4} took 0.472~s.
The configurations of \texttt{model\_3} and of \texttt{model\_4} set 5120 extra MSA rows, and the configuration of \texttt{model\_5} sets 1024 extra MSA rows~\cite{alphafoldrepo}.
The smaller number of extra MSA rows may explain the shorter call of \texttt{model\_5}. We did not test this hypothesis.
Section~\ref{sec:limitations} discusses why times differ between these sweeps and the other experiments.

\subsection{Cost at List Price}\label{sec:res-cost}
Google's on-demand price for a TPU v5e chip in us-west4 was \$1.20 per hour on November 23, 2026~\cite{googletpupricing}.
The course provided the slice, so we did not pay this price.
At \$1.20 per chip-hour, a slice of eight chips costs \$9.60 per hour, whether one chip works or all eight do.
On this slice, 1000 queries of 100 residues on the matched grid would cost \$0.91 with one chip working and \$0.14 with all eight.
At 1000 residues, 1000 queries would cost about \$40 with one chip working and \$5.11 with all eight.
TPU v5e is also offered as slices of one and four chips~\cite{googletpuv5e}.
If one chip rented alone reached the throughput of one chip of our slice, the 1000 queries of 100 residues would cost \$0.11 at the same price per chip.
We did not measure a chip rented alone.
At 1000 residues, the cost on a chip rented alone and the cost on the full slice differ by about 1\%.
Eight chips finish the same queries sooner, but the parallel efficiency is below 100\% at every length.
These costs cover steady-state calls only. They exclude the parameter initialization, the first call at each new length,
the featurization, and the MSA search. They also assume no recycling.
With three recycles, the steady-state call on the fixed sequence took 3.9 times as long as the call without recycling (Section~\ref{sec:res-sweeps}, Table~\ref{tab:sweeps}).

\subsection{Baseline Configuration on a T4 GPU\\ and on a Xeon CPU Runtime}\label{sec:res-platforms}
Table~\ref{tab:platforms} gives the time of the steady-state call on each platform and in each session. On one TPU chip, the steady-state call took 0.470~s.
On the T4, the steady-state call took 13.086~s in the session of August 8, 2026, at 04:50 UTC and a mean of 6.57~s in the session of September 15, 2026, at 19:01 UTC.
On the CPU runtime, the call took 212.113~s in the session of August 8, 2026, at 05:19 UTC and means of 348.9~s to 359.3~s in the three sessions of September 15 to 17, 2026 (Table~\ref{tab:platforms}).
These values come from different sessions and from different copies of the script. We measured the TPU in one period
only. The TPU results were committed between August 8, 2026, at 17:23 UTC and August 11, 2026, at 22:38 UTC.
The times on the T4 and on the CPU runtime changed between sessions. For these reasons, we report only the smallest
ratio over all sessions. We divided the mean steady-state call of each Colab session by the steady-state call on one
TPU chip. Each session of August 8, 2026, timed one call, so the mean of each session is the time of that call.
For the TPU, we used 0.471~s, the longest recorded steady-state call of the 11 single-query runs of the baseline configuration.
In every session, the mean steady-state call on the T4 took more than 13.9 times as long as this call on one TPU chip.
On the CPU runtime, the mean steady-state call took more than 449 times as long.
These minimum ratios allow for the rounding of the recorded TPU time and divide by 0.4715~s, the largest time that rounds to 0.471~s.
Against the 0.470~s of the baseline run on the TPU, both ratios are higher.
Single calls can fall below these minimum ratios.
The fastest single call on the T4 took 6.49~s. The 6.49~s call on the T4 lasted 13.8 times as long as the steady-state call of the baseline run on the TPU (0.470~s).
These ratios also compare different arithmetic. On the TPU, JAX computes float32 matrix products in bfloat16 by
default~\cite{jaxmatmulprecision}.

Our records do not explain the change between sessions. Both T4 sessions report the same GPU model, driver, and
compute capability. The two T4 sessions differ at least in the copy of the script, the install commands, and the NumPy version.
NumPy was 2.5.1 in the T4 session of August 8, 2026, and 2.1.3 in the T4 session of September 15, 2026.
The sessions of August 8, 2026, did not record their JAX version or their host CPU. In public clouds, instances of the same type can
differ in performance, for some providers markedly~\cite{leitner2016patterns}.

\section{Discussion}\label{sec:discussion}

\subsection{Interpretation}
In our runs, the slice did not act as one large device. The model used one chip unless the program mapped work to the
others. This behavior is expected from how JAX places work on devices. Our measurements quantify the effect for
AlphaFold2 on a TPU v5e slice. On a rented slice of eight chips, a program that uses one chip leaves seven chips idle
(Section~\ref{sec:res-cost}).
At 118 residues and up to eight queries per call, batching on one chip with \texttt{vmap} did not replace multi-query \texttt{pmap}.

A user with many independent queries can run one query on each chip in parallel, as multi-query \texttt{pmap} did. Each
\texttt{pmap} call ran queries of one length, so batches of mixed lengths need padding or grouping by length, as
ParaFold does~\cite{zhong2022parafold}.
We did not test the usual alternative to \texttt{pmap}. In this alternative, eight separate processes run at the same time, and each process uses its own chip.
Many users run AlphaFold2 on several accelerators in this way. Our \texttt{pmap} runs used JAX's implementation of
\texttt{pmap} with \texttt{shard\_map}, but we did not call \texttt{jax.shard\_map} directly or shard a batch with
\texttt{jax.jit}.

Automatic partitioning did not shorten the call for one query. Our program gave the partitioner no sharding
annotations, and we found none in the source of AlphaFold2. We did not try to add annotations or to change the model.
The external ensemble ran eight featurizations of one query on eight chips. The external ensemble averaged one output
with a collective operation. With random weights, this output was zero. We did not test whether this average helps
prediction.

\subsection{Limitations}\label{sec:limitations}
The model ran with random weights. The trained weights have the same shapes, so we expect the same steady-state times.
The parameters are an argument of the compiled program, so XLA does not specialize the program on their values. We did
not measure the times with the trained weights. Every MSA held only the query, and no run relaxed the predicted
structure with a force field. We did not measure deep MSAs or the models with templates. Only the recycling sweep ran with recycles. We
timed the calls of the model alone, so our costs cover only part of the work for one query.
We measured one TPU generation, the three models without templates, and inputs of at most 1000 residues.

Most settings ran once. Hoefler and Belli ask authors to report confidence intervals for nondeterministic
measurements~\cite{hoefler2015benchmarking}. Papadopoulos et al. list repeated experiments as the first principle of performance evaluation in the cloud~\cite{papadopoulos2021principles}.
 Most of our settings meet neither of these
recommendations. We measured the TPU in one period only.
In the three repeated runs, the coefficient of variation was at most 1.00\% for every timed call.
The check of the three repeated runs does not cover changes between Jobs or between sessions.

Times also differ between experiments at similar lengths.
At 120 residues, the repeated-alphabet sequence took 0.408~s in \texttt{predict}.
At 118 residues, the fixed sequence took 0.469~s to 0.471~s in \texttt{predict}.
With \texttt{vmap} and $B$ = 1, the repeated-alphabet sequence of 118 residues took 0.4875~s in \texttt{apply}, although \texttt{predict} includes \texttt{apply}.
These runs differ in Job, script, or input family. We did not study these differences. Between experiments, we draw no
conclusion from discrepancies of this size. On the T4 and on the CPU runtime, the changes between sessions were large.

The records have gaps.
The Colab runs of August 8, 2026, used a copy of the script, and this copy exists only inside the notebooks of commit 2755250.
We cannot identify the script of the baseline run on the TPU. The TPU logs are not in the repository. The sweep files are
summaries of the Job output. The committed TPU Jobs pinned only JAX.

\section{Conclusion}\label{sec:conclusion}
AlphaFold2 inference on an eight-chip TPU v5e slice used one chip unless the program mapped the work to the others.
Among the four ways of mapping the work that we compared, only multi-query \texttt{jax.pmap} ran independent queries on all eight chips.
On the matched grid, eight chips gave 6.53 to about 7.9 times the throughput of one chip.
We batched up to eight queries of 118 residues on one chip with \texttt{jax.vmap}, and the throughput did not rise.
Automatic partitioning without sharding annotations did not shorten the call for one query.
For a user with many independent queries, multi-query \texttt{jax.pmap} is a simple way to put every chip of a rented slice to work.
Our measurements used random weights, and most settings ran once on one slice. Future work should repeat the measurements with trained weights.
It should also compare multi-query \texttt{pmap} with eight separate processes running at the same time.

\section*{Data and Code Availability}
The benchmark scripts, the Kubernetes Job definitions, the notebooks, and the result files are in the repository
\url{https://github.com/lorenzopazienza/alphafold-tpu-benchmark}. The code is released under the MIT license. The files
in \texttt{results/sweep/} are summaries of the Job output (Section~\ref{sec:records}).
Commit 2755250 (August 8, 2026, 17:23 UTC) holds the notebooks and most TPU results.
Commit cccd446 (August 8, 2026, 18:05 UTC) holds the matched grid, and commit 625c5dd (August 11, 2026, 22:38 UTC) holds the external ensemble.
The September runs used commits 86fca03 (September 15, 2026, 17:57 UTC) and 9e1a07e (September 16, 2026, 12:35 UTC).
The tag \texttt{arxiv-v1} marks commit a63e955 (September 22, 2026, 13:05 UTC). Versions 1 and 2 cite this commit as the state of the repository.
The tag \texttt{arxiv-v3} marks commit 53f05a7 (November 23, 2026, 21:11 UTC).
The folder \texttt{paper-v3/} at this tag holds the scripts that
generate the tables and the figure of version 3. The scripts that generate the tables and the figure of this
version are in the folder \texttt{paper-v4/} at the tag \texttt{arxiv-v4}.
The result folders of the sessions of September 15 to 17, 2026, store the commit of the script, the package versions, and the host CPU of each session.

\section*{Use of AI Tools}
The first author conceived and directed the study and made all decisions. He launched all runs and checked each result
against the summary files in the repository.
In August 2026, he used Claude (Anthropic) to write the benchmark scripts and the summary files in \texttt{results/sweep/}.
He used Claude Code (Anthropic) to write the later code and analyses and to assist with the writing.
He reviewed the code and the analyses at each step, worked on every sentence of this version by hand, and
is fully responsible for the content of the paper. The second author assisted at times and used Cursor for his
contributions. Each number was checked against the summary file or web page that it cites. A script checks that each printed number equals the rounded value of the source named in the comment next to the number.
The first author checked the pairing of values and settings by hand and with AI assistance, without a dedicated script.
The summary files could not be checked against the output of the TPU Jobs,
which is lost.

\section*{Author Contributions}
L. Pazienza led the study.
He designed it, wrote the benchmark scripts with the assistance of Claude (Use of AI Tools), wrote the Kubernetes Jobs, and ran all the experiments.
 He carried out the re-analysis and wrote this version of the paper without
the second author, with the AI assistance described in Use of AI Tools. I. El Bani helped shape how the research was
framed and presented. He also assisted at
times, built the project website for the course, wrote parts of the first version of the paper, and reviewed all of that
version. He approved the submission of this version.
Commit cccd446, which holds the results of the matched grid, was made by the second author from runs that the first author performed.
The two authors presented the work together at Google headquarters in Mountain View, California, on August 13, 2026.

\section*{Acknowledgment}
The TPU experiments of this work were carried out during the Summer Session 2026 in the Stanford University course ME 344, Introduction to High Performance Computing.
They used TPU~v5e infrastructure that the university provided. We thank Mourad Bouache (Google) and Steve Jones, who taught the course.
Mourad Bouache evaluated our course research project, gave us feedback, and invited us to present the project at Google headquarters in Mountain View, California, on August 13, 2026.
Steve Jones gave us access to the TPU~v5e cluster. He is the Director of the Stanford High Performance Computing Center
and a Research Scientist at Stanford University. We thank the Department of Mechanical Engineering at Stanford
University for hosting the course.

\bibliographystyle{IEEEtran}
\bibliography{references}

\ifappendix
\appendices
\section{Changes from Earlier Versions}\label{app:changes}
This paper is version 4 of arXiv:2609.34818. Version 2 changed only how the abstract on the arXiv page wrote two JAX
function names. The paper itself is the same in versions 1 and 2, so every claim listed here appears in both.
Version 1 had the title \vonequote{Accelerator Choice Is Not Enough: AlphaFold2 Inference on Cloud TPUs.}
Version 3 rewrote the paper. It introduced the current title, shortened the paper from 22 pages to 10, reorganized the sections, and replaced the text of version 1 with a new text based on a re-analysis of the committed result files.
Version 4 corrects version 3 after a further review. We ran no new experiment for versions 3 and 4. We checked the claims and numbers of version 1 against the committed result files. This appendix
lists the claims that changed and the reason for each.

\subsection*{Corrections in Version 4}
The note of Table~\ref{tab:experiments} now lists the runs of the baseline configuration on the TPU. Version 3 wrote that
this configuration ran once in each session.
At 1000 residues, version 3 gave the speedup on eight chips as 7.91 times the throughput of one chip and the parallel efficiency as 98.9\%.
The one-chip throughput at this length has two digits, so this version gives ranges (Section~\ref{sec:res-pmap} and Table~\ref{tab:grid}).
Version 3 printed the coefficients of the power-law fit with two digits. These coefficients do not reproduce the
deviations that version 3 gave, so this version prints four digits.
Version 3 gave the minimum ratio between the CPU runtime and one TPU chip as more than 450.
With the rounding of the TPU time, the bound that holds is 449 (Section~\ref{sec:res-platforms}).
Version 3 claimed that the copy of the parameters could not explain all of the extra time per call.
The extra times, however, are differences of single runs, and the script also copies the inputs inside the timed region (Section~\ref{sec:res-pmap}).
Version 3 did not say that the logits averaged by the external ensemble are zero with
random weights (Section~\ref{sec:res-auto}).
Version 3 gave the throughput of \texttt{vmap} with $B$ = 4 as 0.73 times the throughput of $B$ = 1.
This ratio came from the rounded throughputs. From the recorded times, the ratio is 0.72 (Section~\ref{sec:res-default}).

\subsection*{Ratios Between Platforms (Section~\ref{sec:res-platforms})}
Version 1 reported a ratio of 451 between one TPU chip and the CPU runtime. That version also gave a ratio of 27.8 between one TPU chip and the T4.
Its limitations section said that the Colab values came from \vonequote{the same measurement campaign as the TPU
numbers.}\ Our records do not support this claim.
The T4 and the CPU runtime ran in two separate Colab sessions on August 8, 2026, at 04:50 UTC and 05:19 UTC, respectively.
The baseline run on the TPU did not record its date. The result of this run was committed on August 8, 2026, at 17:23 UTC.
The three values also came from different copies of the script (Table~\ref{tab:platforms}).
In the session of September 15, 2026, at 19:01 UTC, the T4 took a mean of 6.57~s per steady-state call, against 13.086~s in the session of August 8, 2026, at 04:50 UTC.
In the three CPU sessions of September 15 to 17, 2026 (Table~\ref{tab:platforms}), the CPU runtime took means of 348.9~s to 359.3~s, against 212.113~s in the session of August 8, 2026, at 05:19 UTC.
This version gives only the smallest ratio over all sessions (Section~\ref{sec:res-platforms}).
In every session, the mean steady-state call on the T4 took more than 13.9 times as long as the longest recorded steady-state call on one TPU chip (0.471~s).
On the CPU runtime, the mean steady-state call took more than 449 times as long.
Version 1 also gave a ratio between the T4 and the CPU runtime. We removed this ratio.

\subsection*{Speedup on Eight Chips (Section~\ref{sec:res-pmap})}
Version 1 reported that multi-query \texttt{pmap} on eight chips reached 6.92 times the throughput of the baseline run on the TPU, with a parallel efficiency of 86.5\%.
The two runs differed in input and in timed region. This version takes the speedup only from the matched grid. There,
every point used the same script, input family, and timed region.
Eight chips gave 6.53 to about 7.9 times the throughput of one chip.
Version 1 explained the lower parallel efficiency at 100 residues by a fixed overhead per call. This version gives the extra time per call and leaves its cause open.
Version 1 also wrote that \vonequote{distinct per-chip HBM allocations confirm the work is distributed.}\ Memory shows
only that every chip held data. The division of the work follows from \texttt{jax.pmap}, which runs one replica per
device~\cite{jaxpmap}.

\subsection*{First Call and Profiler Teardown (Section~\ref{sec:res-first})}
In every August version that we have, the single-query script runs the first call with the profiler.
The version of commit 86fca03 (September 15, 2026, 17:57 UTC) adds an option that turns the profiler off.
 The TPU runs did not record the duration of the profiler teardown.
In the CPU sessions of September 15, 2026, at 18:04 UTC, September 16, 2026, at 12:37 UTC, and September 17, 2026, at 08:52 UTC, the profiler teardown took 31.7~s to 31.9~s.
In the T4 session of September 15, 2026, at 19:01 UTC, the profiler teardown took 23.7~s.
Version 1 reported that the TPU first call took 59.1 times as long as the steady-state call. That version described this ratio as uncorrected.
This version reports the first call as 27.78~s and gives no ratio.
Table~\ref{tab:platforms} marks that this call probably includes the profiler teardown.
Version 1 also read the compilation-cache experiment as showing that more than half of the first call survives the
cache. Both first calls of that experiment include the profiler teardown, so this version does not attribute the 15.19~s of the second process to compilation.
For the T4 and for the CPU runtime, version 1 subtracted the profiler teardown from the first calls of August 8, 2026.
This version uses the runs of September 15 to 17, 2026, without the profiler instead.

\subsection*{Removed Profiler Trace}
Version 1 reported a trace of one TPU first call.
In the profile of that call, a JAX function that traces and compiles had 12.55~s of self time out of 16.56~s. This share is about 76\%.
We did not keep the trace file, and only a written analysis of it remains. No reader can check the value, so we
removed it.

\subsection*{Automatic Partitioning (Section~\ref{sec:res-auto})}
Version 1 named the automatic-partitioning experiment \vonequote{automatic sharding} and discussed it in terms of GSPMD.
It cited the GSPMD paper for the claim that a partitioner without annotations falls back to replication. That paper
does not make this claim. The runs did not record which partitioner ran. This version reports the experiment as a run
without sharding annotations. It does not say whether the call ran on one chip or on every chip.

\subsection*{Cost Comparison (Section~\ref{sec:res-cost})}
Version 1 found that the slice with one chip working cost about as much per query as the T4.
The values were \$1.25 and \$1.27 per 1000 queries.
This parity held only for the T4 time of the session of August 8, 2026.
The T4 price came from a survey that version 1 did not cite.
The CPU price was our estimate for an instance comparable to the Google Cloud machine type n2-standard-4, with four virtual CPUs.
In the sessions of September 15 to 17, 2026, the CPU runtime had two virtual CPUs, and the sessions of August 8, 2026, did not record this number.
Version 1 also priced hypothetical slices of one, two, and four chips in a table. This version prices only the TPU, at Google's list rate of \$1.20 per chip-hour.
It gives one conditional example for a chip rented alone.

\subsection*{Removed \AFthree{} Results}
Version 1 had a section and an appendix on \AFthree{}. The release we tested did not accept the TPU as a platform.
The \AFthree{} runs do not address the question of this paper. That question concerns the TPU slice. We removed the
section, the appendix, and all the \AFthree{} numbers.

\subsection*{Other Corrections}
Version 1 stated that the power-law fit used a sequence-length sweep from 100 to 1000 residues. The fit in fact uses the matched grid.
The sequence-length sweep covered 60 to 500 residues.
Version 1 wrote that none of the cited work reported TPU experiments, but that version cited a model trained on TPUs.
Version 1 described a controlled comparison with identical code on the three platforms.
AlphaFold2's source ran unmodified, but the benchmark scripts differed between the platforms and between the sessions (Table~\ref{tab:platforms}).
Version 1 called the copy of the script in the notebooks of August 8, 2026, uncommitted.

The notebooks in commit 2755250 hold this copy.
Seven works cited in version 1 do not make the statements attributed to them in that version. We corrected these
statements or removed the citations. Version 1 treated the files in \texttt{results/sweep/} as raw results. These files
are summaries of the Job output (Section~\ref{sec:records}). The abstract now states that the weights were random. We
also renamed three terms. A \vonequote{pod} is now a slice, a \vonequote{protein} is a query, and \vonequote{automatic
sharding} is automatic partitioning.

\fi

\end{document}